\documentclass[12pt]{article}
\usepackage{amssymb,amsmath,amscd,amsthm}
\usepackage[cp1251]{inputenc}
\usepackage[T2A]{fontenc}
\usepackage{pstricks}
\usepackage{graphicx,psfrag}

\usepackage[english]{babel}

\newtheorem{theorem}{Theorem}[section]

\newtheorem{lemma}[theorem]{Lemma}

\date{}

\begin{document}

\title{Applications of elliptic operator theory to the isotropic interior transmission eigenvalue problem}

\author{ E.Lakshtanov\thanks{Department of Mathematics, Aveiro University, Aveiro 3810, Portugal.   This work was supported by {\it FEDER} funds through {\it COMPETE}--Operational Programme Factors of Competitiveness (``Programa Operacional Factores de Competitividade'') and by Portuguese funds through the {\it Center for Research and Development in Mathematics and Applications} (University of Aveiro) and the Portuguese Foundation for Science and Technology (``FCT--Fund\c{c}\~{a}o para a Ci\^{e}ncia e a Tecnologia''), within project PEst-C/MAT/UI4106/2011 with COMPETE number FCOMP-01-0124-FEDER-022690, and by the FCT research project
PTDC/MAT/113470/2009 (lakshtanov@rambler.ru).} \and
B.Vainberg\thanks{Department
of Mathematics and Statistics, University of North Carolina,
Charlotte, NC 28223, USA. The work was partially supported   by the NSF grant DMS-1008132 (brvainbe@uncc.edu).}}

\maketitle

\begin{abstract}
The paper concerns the isotropic interior transmission eigenvalue (ITE) problem.
This problem is not elliptic, but we show that, using the Dirichlet-to-Neumann map, it can be reduced to an elliptic one.
This leads to the discreteness of the spectrum as well as to certain results on possible location  of the transmission eigenvalues.
If the index of refraction $\sqrt{n(x)}$ is real, we get a result on the existence of infinitely many positive ITEs
and the Weyl type lower bound on its counting function. All the results are obtained under the assumption that $n(x)-1$ does not vanish at the boundary of the obstacle or it vanishes identically, but its normal derivative does not vanish at the boundary. We consider the classical transmission problem as well as the case when the inhomogeneous medium contains an obstacle. Some results on the discreteness and localization of the spectrum are obtained for complex valued $n(x)$.
\end{abstract}

\textbf{Key words:}
Isotropic interior transmission eigenvalue, parameter-elliptic problem, counting function,  Weyl formula.

\section{Introduction.}

Let us recall that $\lambda   \in \mathbb C$ is called an \textit{interior transmission eigenvalue} (ITE) if
the homogeneous problem
\begin{equation}\label{Anone0}
-\Delta u - \lambda u =0, \quad x \in \mathcal O, \quad u\in H^2(\mathcal O),
\end{equation}
\begin{equation}\label{Anone}
-\Delta v - \lambda   n(x)v =0, \quad x \in \mathcal O, \quad v\in H^2(\mathcal O),
\end{equation}
\begin{equation}\label{Antwo}
\begin{array}{l}
u-v=0, \quad x \in \partial \mathcal O, \\
\frac{\partial u}{\partial \nu} - \frac{\partial v}{\partial \nu}=0, \quad x \in \partial \mathcal O,
\end{array}
\end{equation}
has a non-trivial solution.
Here $\mathcal O\subset \mathbb R^d$ is a bounded domain with a $C^\infty$-boundary, $H^{2}(\mathcal O), ~H^{s}(\partial \mathcal O)$ are the Sobolev spaces, $n(x) \! \neq \! 0, ~ \! x \! \in \! \overline{\mathcal O}$ is  a complex $C^\infty$-function, $\nu$ is the outward unit normal vector.

Problem (\ref{Anone0})-(\ref{Antwo}) appears naturally when the scattering of plane waves is considered, and
the inhomogeneity in $\mathbb R^d$ is located in $\mathcal O$ and is described by  the
index of refraction $\sqrt{n}$.  We will be concerned with the cases $d=2,3$.
Infinite differentiability of $\partial \mathcal O$ and $n$ is assumed for the sake
of simplicity, a finite smoothness is enough for all the results below. There are weaker definitions of ITE-s when solutions $(u,v)$ of the problem (\ref{Anone0})-(\ref{Antwo}) are assumed to be only square integrable (boundary conditions (\ref{Antwo}) still can be defined since $u$ and $v$ satisfy the homogeneous elliptic equations). It will be shown in the Attachment that these weak eigenfunctions of the ITE problem belong to the Sobolev space $H^2$ under conditions imposed in the present paper, i.e., the a priory assumption $u,v\in H^2(\mathcal O)$ does not reduce the set of ITE-s.

We also consider the case when $\mathcal O$ contains a compact obstacle $\mathcal V \subset \mathcal O$, $\partial \mathcal V \in C^\infty$.
In this case, equation (\ref{Anone}) is replaced by
\begin{equation}\label{AnoneB}
-\Delta v - \lambda   n(x)v =0, \quad x \in \mathcal O \backslash \mathcal V, \quad v\in H^2(\mathcal O \backslash \mathcal V);~~~
v(x)=0, \quad x \in \partial \mathcal V,
\end{equation}
while equation (\ref{Anone0}) remains valid in $\mathcal O$. For simplicity of notations, we will consider problem (\ref{Anone0})-(\ref{Antwo}) as a particular case of (\ref{Anone0}),(\ref{AnoneB}),(\ref{Antwo}) with $\mathcal V = \emptyset$. The Dirichlet boundary condition on $\partial\mathcal V$ in (\ref{AnoneB}) (as well as in our previous papers on ITEs) can be replaced by the Neumann or Robin boundary condition without any changes in the results or proofs.

Note that problem (\ref{Anone0})-(\ref{Antwo}) is neither elliptic, nor formally-symmetric, and therefore the properties of its spectrum can not be obtained by soft arguments. Discreteness of the ITEs was proved first in \cite{CKP} in the case when $n$ is real and $n(x)-1$ preserves the sign in the whole domain $\overline{\mathcal O}$. This result was extended in \cite{CCH} to the case of domains with cavities, i.e., $n(x)\!-\!1$ was allowed to vanish inside $\mathcal O$. In \cite{sylv}, it was proved that the set of the ITEs is discrete if $\mathcal V = \emptyset$ and $n(x)-1$ does not vanish  at the boundary $\partial \mathcal O$ ($n$ can be complex valued).
The case of piece-wise constant $n$ was studied in \cite{HC}. In particular, it was shown there that the negative semi-axis  does
not contain ITEs when $n\neq 1$ is a constant.

The index of refraction $\sqrt{n}$ is assumed to be real valued when the positive ITEs are studied. In \cite{Dr} it was proved that the set of the positive ITEs is infinite if $\mathcal V = \emptyset$, and the function $n(x)\!-\!1$ is not zero for all $x\in \overline{\mathcal O}$. Some Weyl type lower estimates on the counting function for the positive ITEs were obtained in \cite{ss} in the case when $n\!>\!1$ everywhere inside $\mathcal O$ (see also \cite{tsm}). Note that $n(x)=1$ at the boundary was allowed in \cite{ss}, and the discreteness of the spectrum in this situation was justified there.
The discreteness of ITEs and the existence of infinitely many positive ITEs was proved in \cite{chobst} when  $\mathcal V \neq \emptyset$ and $n<1$ everywhere.

In \cite{tsm} it is shown that in the case of $n(x)>1, x \in \overline{\mathcal O},$  all but finitely many complex transmission eigenvalues are confined to a parabolic neighborhood of the real positive semi-axis. In \cite{tsm2} the authors justified the completeness of the set of the interior transmission eigenfunctions under the same assumption on $n(x)$.

In \cite{LV4},\cite{lakvain5},\cite{lakvain6}, we considered anisotropic problems and proved the discreteness of the ITEs, the existence of real ITEs, and established the Weyl type estimates for the real ITEs. Some of these results were known earlier (see the recent review \cite{HadCak}).  We showed that, under weak assumptions, the anisotropic ITE problem is parameter-elliptic. This allowed us to broaden the scope of applications, simplify the proofs and obtain new results. However, our approach can not be directly applied to the isotropic case that we consider here, since the isotropic problem is not elliptic. The extension of our previous results to the isotropic problems will be obtained in the present paper by the reduction of the problem to an elliptic pseudo-differential operator (of a lower order) at the boundary using Dirichlet-to-Neumann maps for equations (\ref{Anone0}) and (\ref{AnoneB}).

The isotropic problem  (\ref{Anone0}),(\ref{AnoneB}),(\ref{Antwo}) is considered in this paper when
either
\begin{equation}\label{grcond}
n(x)-1 \neq 0, \quad x \in \partial \mathcal O,
\end{equation}
or
\begin{equation}\label{grcond1}
n(x)-1 \equiv 0, \quad \frac{\partial}{\partial \nu} n(x) \neq 0, \quad x \in \partial \mathcal O.
\end{equation}
Let us stress again that these conditions on $n$ are imposed only at the boundary of the domain. Condition (\ref{grcond1}) allows one to consider scattering problems in inhomogeneous
media with continuous index of refraction $\sqrt{n(x)}$.  We also assume that there exists a closed sector  $\Lambda \subset \mathbb  C$  centered in the origin that does not contain any points of the following set $\mathcal N$:
\begin{equation}\label{en}
\mathcal N = \{1\} \cup \left \{\frac{1}{n(x)}, ~ x \in \overline{ \mathcal O} \right \}.
\end{equation}
If function $n(x)>0,x \in \overline{ \mathcal O},$  is real-valued, then the latter assumption obviously holds for any sector that does not contain the positive semi-axis $\mathbb R^+$.

We will show that the set of ITEs is discrete with two possible accumulation points: zero and infinity, and zero is not an accumulation point if $\mathcal V = \emptyset$. Moreover, if $\mathcal V = \emptyset$ we show that there are at most finitely many ITEs in any closed sector $\Lambda \subset \mathbb  C$  centered in the origin
that does not contain any points of $\mathcal N$. The same is true if $\mathcal V \neq \emptyset$ and a neighborhood of the origin is cut off from $\Lambda $, i.e., $\Lambda $ is replaced by $\Lambda \bigcap \{|\lambda|>1\}$.


We will also prove the existence of infinitely many \textit{real} ITEs when $n(x)\!>\!0, ~\! x \!\in \!\overline{\mathcal O},$ is a real-valued function and
$$
\sigma \left ( Vol(\mathcal O) - \int_{\mathcal O \backslash \mathcal V} n^{d/2}(x)dx \right )>0,
$$
where
\begin{equation}\label{dsig}
\sigma={\rm sign}(n(x)-1),~~x\in \mathcal O,~~~\text{dist}(x,\partial \mathcal O)\ll 1,
 \end{equation}
 is the sign of $n\!\!-\!\!1$ in a neighborhood of $\partial\Omega$ strictly inside of $\Omega$. The constant $\sigma$ is well defined due to the conditions imposed on $n$. Moreover, we will obtain a Weyl type lower bound on the counting function $N_T(\lambda)$ of the positive ITEs. Note that the condition imposed above holds, for example, in the following cases: 1) $n(x)>1$ on $\partial \mathcal O$, but $n(x)$ is small enough inside of $ \mathcal O$, or 2) $n(x)>1$ everywhere, but the obstacle $\mathcal V$ is large enough, or 3) $n(x)<1$ on $\partial \mathcal O$, but $n(x)$ is large enough inside of $ \mathcal O$.

All the proofs are based on methods of elliptic pseudo-differential operators (p.d.o). Therefore the assumptions on the smoothness
of $\partial \mathcal O$
and $n(x)$ are essential for us, while some of the earlier results mentioned above were proved for $n(x) \in L^\infty$. Let us outline how methods of elliptic equations appear in the study of non-elliptic problem (\ref{Anone0}),(\ref{Antwo}),(\ref{AnoneB}).

 Let
\begin{equation}\label{ffa}
F(\lambda), F_n(\lambda):H^{3/2}(\partial\mathcal O)\to H^{1/2}(\partial\mathcal O)
\end{equation}
be the Dirichlet-to-Neumann map for equations (\ref{Anone0}) and (\ref{AnoneB}), respectively. Note, that the domain of the operator $F_n$ consists of functions defined on $\partial\mathcal O$, since the value $v=0$ is fixed on $\partial\mathcal V$ (if $\mathcal V\neq \emptyset)$.
Operators (\ref{ffa}) are well defined when $\lambda$ is not an eigenvalue of the Dirichlet problem for equations (\ref{Anone0}) and (\ref{AnoneB}), respectively. In
particular, from the Green formula, it immediately follows that these operators are well defined in any sector $\Lambda$ that does not contains points of  set
$\mathcal N$.

Consider an arbitrary $\lambda=\lambda_0$ that is neither a pole of $F(\lambda_0)$ nor $F_n(\lambda_0)$. From (\ref{Antwo}) it follows that $\lambda=\lambda_0$ is an ITE if and only if the kernel of $F_n(\lambda_0)-F(\lambda_0)$ is not empty. The situation when  $\lambda=\lambda_0$ is a pole is quite similar and will be considered later.
Operators (\ref{ffa}) are elliptic p.d.o. of the first order. Their principal symbols do not depend on $n$, i.e., the principal symbol of the difference $F_n\!-\!F$ is zero. However, this difference can be an elliptic operator of a lower order.

Our results are based essentially on old papers by B.Vainberg and V. Grushin \cite{VGI},\cite{VG2}, who calculated the full symbols of different pseudo-differential operators that map boundary values of one problem for an elliptic equation to the boundary values of another problem (their goal was to show that coercivity in some non-elliptic problems may occur due to the structure of the lower order terms of the full symbol of the operator). We will apply these calculations to prove the following two main lemmas.
\begin{lemma}\label{lemmasigma}
 Let one of the conditions (\ref{grcond}) or (\ref{grcond1}) hold. Let $s=1$ if (\ref{grcond}) holds, and $s=2$ if (\ref{grcond1}) holds. Then the following statements
 are valid

1)  The difference
\begin{equation}\label{diff}
F_n(\lambda)-F(\lambda):H^{\frac{3}{2}}(\partial\mathcal O)\to H^{\frac{3}{2}+s}(\partial\mathcal O),~~ \lambda \neq 0,
\end{equation}
 depends meromorphically on~$\lambda$ and is an elliptic pseudo-differential operator of order $-s$, when $\lambda$ is not a pole of (\ref{diff}).

 If $\lambda$ is not a pole of the operator $F_n-F$, and $|\xi^*|$ is the length of the covector defined in (\ref{xis}), then the  principal symbol of $F_n(\lambda)-F(\lambda)$ is
\begin{eqnarray}\label{Grcond3}
(1-n(x))\frac{\lambda }{2|\xi^*|}, \quad x \in \partial \mathcal O, \quad \xi \in \mathbb R^{d-1}, ~~\text{if (\ref{grcond}) holds}, \\
\frac{1}{4}\frac{\partial n(x)}{\partial \nu}\frac{\lambda }{|\xi^*|^{2}}, \quad x \in \partial \mathcal O, \quad \xi \in \mathbb R^{d-1}, ~~\text{if (\ref{grcond1}) holds}\label{grcond3a}.
\end{eqnarray}

Each pole $\lambda=\lambda_0$ of this operator has order one, and the residue $P_{\lambda_0}$ has infinitely smooth
integral kernel $P_{\lambda_0}(x,y)\in C^\infty$ in $(x,y)$. Operator  $P_{\lambda_0}$ is a projection on a finite dimensional space spanned by the normal derivatives of the solutions of the homogeneous Dirichlet problems for equations (\ref{Anone0}) and (\ref{AnoneB}). The regular part $F_n(\lambda)-F(\lambda)-\frac{P_{\lambda_0}}{\lambda-\lambda_0}$ of operator (\ref{diff}) is a p.d.o. of order $-s$, whose principal symbol is given by (\ref{Grcond3}),(\ref{grcond3a}).

2) If $\mathcal V = \emptyset$, then $F_n(0)=F(0) $, and the operator
$$
G(\lambda):=\frac{F_n(\lambda)-F(\lambda)}{\lambda}:H^{\frac{3}{2}}(\partial\mathcal O)\to H^{\frac{3}{2}+s}(\partial\mathcal O),~~ |\lambda |\ll 1,
$$
has a limiting value $G(0)$. Operator $G(\lambda)$ depends analytically on $\lambda$ in a neighborhood of the origin and is Fredholm for each $\lambda$.
\end{lemma}

\noindent
\textbf{Remark 1.} The upper index $3/2$ in (\ref{diff}) can be replaced by any $m\in \mathbb R.$ We decided to use $m=3/2$, since a part of the proof is slightly simpler in this case.

\noindent
\textbf{Remark 2.} Operators $F_n$ and $F$ have order one, and the difference $F_n-F$ has order $-1$ or $-2$ (it is equal to $-s$), i.e., two or three first terms of the full symbol are canceled when the difference is taken.

The second lemma concerns the parameter-ellipticity of operator $\lambda^{-1} [F_n(\lambda)-F(\lambda)]$, which leads to the
invertibility of this operator for large $|\lambda|$. We will formulate here the invertibility result.

Let $H^{m,k}(\partial \mathcal O),~k>0,$ be the Hilbert space with the norm
\begin{equation}\label{norm}
||u||^2_{H^{m,k}(\partial \mathcal O)}=||u||^2_{H^{m}(\partial \mathcal O)}+k^{2m}||u||^2_{L^2(\partial \mathcal O)},
\end{equation}
where $H^{n}(\partial \mathcal O)$ is the Sobolev space. Below, parameter $k$ will be always equal to $\sqrt{|\lambda|}$.
Let  $\Lambda$ be an arbitrary closed sector of the complex plane that does not contain the set $\mathcal N$
defined in (\ref{en}), and let $\Lambda'=\Lambda\bigcap\{|\lambda|>1\}$. Recall that $\Lambda$ does not contain poles of $F_n$ and $F$.
\begin{lemma}\label{paramellTh}
Let $n(x)\neq 1$ on $\partial \mathcal O$ (i.e. (\ref{grcond}) holds) and let sector $\Lambda$ not contain any points of $\mathcal N$. Then for each $\Lambda'$ and each $m\in \mathbb R$, the operator
\begin{equation}\label{s1}
\lambda^{-1} [F_n(\lambda)-F(\lambda)]:H^{m,k}(\partial \mathcal O)\to H^{m+1,k}(\partial \mathcal O),\quad  \lambda\in\Lambda',\quad k=\sqrt{|\lambda|},
\end{equation}
is uniformly bounded  in $\lambda$. Moreover,  there exists $A=A(\Lambda')$ such that operator (\ref{s1}) is invertible when $|k| >A$ and
\begin{equation}\label{s2}
\parallel (F_n(\lambda)-F(\lambda))^{-1}f\parallel_{H^{m,k}}\leq C|\lambda|^{-1}\parallel f\parallel_{H^{m+1,k}}, \quad \lambda\in\Lambda',
\quad |k| > A.
\end{equation}
Let $n(x)\equiv 1, \frac{\partial n}{\partial \nu}\neq 1$ on $\partial \mathcal O$ (i.e. (\ref{grcond1}) holds) and let sector $\Lambda$ not contain any points of $\mathcal N$. Then for each $\Lambda'$  and each $m\in \mathbb R$, the operator
\begin{equation}\label{s3}
\lambda ^{-1}[F_n(\lambda)-F(\lambda)]:H^{m,k}(\partial \mathcal O)\to H^{m+2,k}(\partial \mathcal O),\quad  \lambda\in\Lambda',\quad k=\sqrt{|\lambda|},
\end{equation}
is uniformly bounded  in $\lambda$. Moreover,  there exists $A=A(\Lambda')$ such that operator (\ref{s3}) is invertible when $|k| >A$ and
\begin{equation}\label{s4}
\parallel (F_n(\lambda)-F(\lambda))^{-1}f\parallel_{H^{m,k}}\leq C|\lambda|^{-1}\parallel f\parallel_{H^{m+2,k}}, \quad \lambda\in\Lambda',\quad |k|>A.
\end{equation}
\end{lemma}

Lemmas \ref{lemmasigma},\ref{paramellTh} imply the following statement.

\begin{theorem}\label{theoremObstacle}
Let one of the conditions (\ref{grcond}) or (\ref{grcond1}) hold. Assume that there exists a closed sector  $\Lambda \subset \mathbb  C$  centered in the origin that does not contain any points of the set $\mathcal N$. Then the following statements hold.

1) If $\mathcal V = \emptyset$, then the set of the ITEs for problem (\ref{Anone0}),(\ref{AnoneB}),(\ref{Antwo}) is discrete with the only possible accumulation point at infinity. Moreover, there is at most a finite number of the ITEs inside each closed sector $\Lambda$ centered in the origin that does not contain points of $\mathcal N$. Thus, if $n(x)$ is real valued  at the boundary, then there is at most a finite number of the ITEs inside each closed sector $\Lambda$ of complex $\lambda$-plane that
does not contain the ray $\mathbb R^+$.

2) If $\mathcal V \neq \emptyset$, then the set of the ITEs for problem (\ref{Anone0}),(\ref{AnoneB}),(\ref{Antwo}) is discrete with the only possible accumulation points at zero and infinity. Moreover, there is at most a finite number of the ITEs inside $\Lambda'=\Lambda\bigcap \{|\lambda|>1\}$.
\end{theorem}
\noindent
\textbf{Remark 3.} If $\mathcal V = \emptyset$, then $F_n(0)=F(0)$, and therefore $\lambda=0$ is an ITE of infinite multiplicity. The multiplicities of all the other ITEs are finite due to lemma \ref{lemmasigma}. If $\mathcal V \neq \emptyset$, then $\lambda=0$ is not an ITE. The latter can be proved very easily for domains with Lipshitz boundary and without assumptions on the smoothness of $n$. Indeed, assume that there exists  a solution $(u,v)$ of (\ref{Anone0}),(\ref{Antwo}),(\ref{AnoneB})  for $\lambda=0$.
Denote $w=u-v \in H^2(\mathcal O \backslash \mathcal V)$. Since $w =\frac{\partial w}{\partial \nu}\equiv 0$ on $\partial \mathcal O$, we have $w \equiv 0$ in $\mathcal O \backslash \mathcal V$. Thus $u \equiv v \equiv 0$ on the boundary $\partial \mathcal V$. Therefore $u$ equals zero in $\mathcal V$, and therefore $u  \equiv 0$ everywhere in $\mathcal O$.

\noindent
\textbf{Remark 4.} The results above can be easily extended to the case when
$$
\frac{\partial^i (n(x)-1)}{\partial \nu^i} \equiv  0, ~  i=0 \ldots m-1, ~ \frac{\partial^m (n(x)-1)}{\partial \nu^m} \neq 0,  \quad x \in \partial \mathcal O.
$$
\noindent

 The following theorem is based on Lemmas \ref{lemmasigma},\ref{paramellTh} and the ideas developed in \cite{lakvain6}.

\begin{theorem}\label{th12} Let $n(x),x\in \mathcal O,$ be a real valued function and let one of the assumptions (\ref{grcond}) or (\ref{grcond1}) hold.
Let
$$
\gamma := \sigma \left ( Vol(\mathcal O) -\int_{\mathcal O \backslash \mathcal V} n^{d/2}(x)dx \right )>0,
$$
where $\sigma$ is defined in (\ref{dsig}).

Then the set of positive ITEs is infinite, and moreover
$$
N_T(\lambda)   \geq \frac{\omega_d}{(2\pi)^d} \gamma   \lambda^{d/2} + O(\lambda^{(d-1)/2}), \quad \lambda \rightarrow \infty,
$$
where $\omega_d$ is the volume of the unit ball in $\mathbb R^d$ and $N_T$ is the counting function of the positive ITEs.
\end{theorem}

Now we will prove Theorems \ref{theoremObstacle} and \ref{th12}. In Section~\ref{mmm}, the results from \cite{VG2} needed to prove the main lemmas will be reviewed. The lemmas will be proved in the last section.

\section{Proofs of the main theorems.}

\textbf{Proof of Theorem \ref{theoremObstacle}.} Obviously, every $\lambda=\lambda_0$ that is not an eigenvalue of the Dirichlet problem for equation (\ref{Anone0}) or (\ref{AnoneB}) is an ITE if and only if the kernel of $F_n(\lambda_0)-F(\lambda_0)$ is not empty. Thus, Theorem  \ref{theoremObstacle} will be proved if we justify the corresponding statements for the set of $\lambda_i$ with non-empty kernel of $F_n(\lambda_i)-F(\lambda_i)$ instead of the set of the ITEs.

We will need the following extension \cite{bleher} of the analytic Fredholm theorem: if a domain $\Omega\in \mathbb{C}$ is connected, and an operator function $T=T(\lambda),~\lambda\in \Omega,$ is  finitely-meromorphic and Fredholm, then the invertibility of $T(\lambda)$ at one point $\lambda\in \Omega$ implies that the inverse operator function $T^{-1}(\lambda),~\lambda\in \Omega,$ is finitely-meromorphic and Fredholm. Recall that a  meromorphic operator function $T=T(\lambda): H_1\to H_2$ in Hilbert spaces $H_i$ is called finitely-meromorphic if the principal part of the Laurent expansion at each pole $\lambda=\lambda_0$ is an operator of a finite rank (i.e., coefficients for negative powers of $\lambda-\lambda_0$ are finite-dimensional operators). This operator function is called Fredholm if operator $T(\lambda)$ is Fredholm at each regular point  $\lambda=\lambda_0$, and the regular part of $T(\lambda)$ is Fredholm at each pole of the function.

The first part of Lemma \ref{lemmasigma} implies that the family of operators (\ref{diff}) is finitely-meromorphic and Fredholm when $\lambda\in \mathbb{C}\backslash \ 0$. Lemma \ref{paramellTh} guarantees  the invertibility of  (\ref{diff}) when $\lambda\in\Lambda\bigcap \{|\lambda|>a\}$ and $a=a(\Lambda)$ is large enough. Thus, the above theorem on the inversion of the meromorphic family of operators can be applied, which leads to the discreteness of the ITEs  in $\mathbb{C}\backslash \ 0$. In order to complete the proof of Theorem \ref{theoremObstacle}, it remains only to show that $\lambda=0$ can not be a limiting point for the set of the ITEs if $\mathcal V = \emptyset$.

Assume that $\mathcal V = \emptyset$. Since operator  (\ref{diff}) is invertible except possibly for a discrete set of points $\lambda$, the same is true for the operator $G(\lambda)=\lambda^{-1}[F_n-F]$.
This and the second statement of Lemma \ref{lemmasigma} allow us to apply the analytic Fredholm theorem to $G(\lambda)$ in a neighborhood of the origin. Hence $G(\lambda)$ and therefore $F_n-F=\lambda G(\lambda)$ may have non-trivial kernels at most at finitely many points of this neighborhood.

\qed

\noindent
\textbf{Proof of Theorem \ref{th12}.} Recall that $n(x), x \in \mathcal O,$ in the statement of the theorem is assumed to be a real valued function .

Denote the set of the {\it positive} ITEs with their multiplicities taken into account by $\{\lambda^T_i\}$. Similarly, denote the set of positive eigenvalues of the Dirichlet problem for $-\Delta$ in $\mathcal O$ by $\{\lambda_i\}$, and the set of positive $\lambda>0$ for which equation (\ref{AnoneB}) in $\mathcal O \backslash \mathcal V$ with the Dirichlet boundary condition at the boundary
$\partial\left(\mathcal O \backslash   \mathcal V\right)$ has a nontrivial solution by $\{ \lambda^n_i\}$.
The corresponding counting functions will be denoted by
\begin{equation}\label{countdef}
N_T(\lambda)=    \#\{i:\alpha<\lambda^T_i \leq \lambda\} , \quad N(\lambda)=     \#\{i:\lambda_i \leq \lambda\}, \quad N_n(\lambda)=     \#\{i:\lambda^n_i \leq \lambda\},
\end{equation}
where $\alpha\in(0,\min(\lambda_0,\lambda^n_0))$ is an arbitrary small enough positive number that does not belong to the set $\{\lambda^T_i\}$. Note that we do not count positive ITEs on the segment $[0,\alpha]$ where we can not guarantee that the number of ITEs is finite if $\mathcal V\neq \emptyset$.

We are going to prove the following estimate on $N_T(\lambda)$ from below through the counting functions $N(\lambda),N_n(\lambda)$:
\begin{theorem}\label{lemmamain1} Let one of the assumptions (\ref{grcond}) or (\ref{grcond1}) hold. Then there exists a constant  $n^-(1)\geq 0$ such that
\begin{equation}\label{mainIneq}
N_T(\lambda) \geq \sigma(N(\lambda)-N_n(\lambda))-n^-(\alpha), \quad \lambda >\alpha,
\end{equation}
where $\sigma$ is defined in (\ref{dsig}).
\end{theorem}
\noindent
\textbf{Remark.} The constant $n^-(\alpha)$ will be defined below during the proof.
\\

Theorem \ref{th12} is a direct consequence of Theorem \ref{lemmamain1} and the well-known (e.g., \cite[Th. 1.2.1]{safvas}) Weyl formula for $N(\lambda)$ and $N_n(\lambda)$. Hence we need only to prove Theorem \ref{lemmamain1}
\\

\noindent
\textbf{Proof of Theorem \ref{lemmamain1}.} An analogue of Theorem \ref{lemmamain1} (and of Theorem \ref{th12}) for anisotropic media was proved in
\cite{lakvain6}. The proof was based on the ellipticity of the operator $F_A-F $, where $A$ is the matrix that describes the anisotropy of the medium. Lemma \ref{lemmasigma} allows us to carry over all the arguments to the isotropic case. One of the main differences is that operator (\ref{diff}) is of negative order (the principal symbols of the terms in the difference get canceled), while the order of $F_A-F $ is positive. Secondly, we will need to be more careful around the point $\lambda=0$ where the ellipticity of (\ref{diff}) is lost. The proof of Theorem \ref{lemmamain1} will be close to the one from \cite{lakvain6}. We will provide a detailed proof below not only because of the differences mentioned above, but also because we were able to simplify some of the steps from \cite{lakvain6}.

We will assume first that
\begin{equation}\label{28111}
\{ \lambda_i\} \cap \{ \lambda^n_i\} = \emptyset.
\end{equation}
This case is more transparent. All additional details, needed to consider the general case (when (\ref{28111}) is violated), will be discussed at the very end of the proof.

\textit{Step 1. Operator $B(\lambda)$ and its eigenvalues $\mu_j=\mu_j(\lambda)$}. We will say that a meromorphic operator function has \textit{a kernel at a pole }(of the first order) if there is a non empty intersection of the kernel of its residue with the kernel of its regular part. The dimension of this intersection will be called the dimension of the kernel  of the operator.

From the definition of the ITEs and (\ref{28111}) it follows immediately that a point $\lambda=\lambda_0\in \mathbb R$ is an ITE if and only if the operator $F(\lambda) - F_n(\lambda)$ has a non-trivial kernel at $\lambda=\lambda_0$. The multiplicity of the ITE coincides with the dimension of the kernel.

We will assume that (\ref{grcond}) holds. Consider the operator
\begin{equation}\label{bla}
 B(\lambda):=\sigma D (F(\lambda)-F_n(\lambda)) D ~ \! : \! ~ H^{3/2}(\partial\mathcal O) \rightarrow H^{1/2}(\partial\mathcal O), \quad D=(1-\Delta_{\partial\mathcal O})^{1/2},
\end{equation}
 where $\Delta_{\partial\mathcal O}$ is the Laplace-Beltrami operator on $\partial\mathcal O$. If condition  (\ref{grcond1}) holds instead of (\ref{grcond}), then one needs only to replace $D$ above by $D^{3/2}$  (alternatively, one can use $D^2$ and replace $H^{1/2}$ by $H^{-1/2}$). The dimensions of the kernels of operators $B(\lambda)$ and $F(\lambda)-F_n(\lambda)$ coincide, and therefore the following lemma is valid.

\begin{lemma}\label{lemma11}
Let (\ref{28111}) hold. Then $\lambda=\lambda_0$ is an ITE if and only if the operator $ B(\lambda)$ has a non-empty kernel at $\lambda=\lambda_0$. The multiplicity of the ITE $\lambda_0$ is equal to the dimension of the kernel of $ B(\lambda_0)$.
\end{lemma}

 We will use the operator $B(\lambda)$ to count the number of the ITEs with their multiplicities taken into account. For this purpose, we are going to study the negative spectrum of the operator $B(\lambda)$.

From the Green formulas for equations (\ref{Anone0}) and (\ref{AnoneB}), it follows immediately that operators $F_n$ and $F$ (and therefore, $B(\lambda)$) are symmetric when
 $\lambda$ is real. By lemma \ref{lemmasigma}, operator $B(\lambda)$ is an elliptic p.d.o. of order one. Hence, if $\lambda$ is not a pole of $B$, then the spectrum of  $B(\lambda)$ consists of a sequence $\{\mu_j(\lambda)\}$ of real eigenvalues of finite multiplicities, and
\begin{equation}\label{minf}
|\mu_j(\lambda)|\to \infty \quad \text{as} \quad  j\to \infty.
\end{equation}
Note that the reason for introducing the operator $D$ in (\ref{bla}) (which was not used in \cite{lakvain6}) is to avoid considering the essential spectrum of the compact operator $F(\lambda)-F_n(\lambda)$ at the point $\mu=0$.

The operator $B=\sigma D (F(\lambda)-F_n(\lambda)) D$ has a positive principal symbol (see (\ref{Grcond3}), (\ref{grcond3a}), and (\ref{dsig})), and therefore (see \cite[Cor. 9.3]{shubin}) it is bounded from below when $\lambda$ is not a pole. Obviously, the bound can be chosen locally uniformly in $\lambda$, i.e., the following statement holds.
\begin{lemma}\label{lemmaShubin}
For each closed interval $I \subset \mathbb R^+$ where the operator $B(\lambda)$ is analytic, there exists a constant $C=C(I)$ such that
\[
\mu_j(\lambda)\geq -C, \quad   \lambda\in I, ~~j=1,2,... ~.
\]
\end{lemma}

\begin{lemma}\label{hardDayLemma}
If operator $B(\lambda)$  is analytic in a neighborhood of a point $\lambda=\lambda_0$, then all the eigenvalues $ \mu=\mu_j(\lambda)$ are analytic in this neighborhood.

If  $\lambda=\lambda_0$ is a pole of the first order of the operator  $B(\lambda)$ and $p$ is the rank of its residue $P$, then $p$ eigenvalues  $ \mu=\mu_j(\lambda)$ and their eigenfunctions have a pole at
$\lambda_0$ and all the others are analytic in this neighborhood. The residues of the eigenvalues  $ \mu_j(\lambda)$ are the eigenvalues of the residue $P$ of the operator $B(\lambda)$.
\end{lemma}
\noindent
{\bf Proof.} The first statement is a well-known property of analytic self adjoint operators (see \cite[Th. XII.13]{reedsimon}) whose spectrum consists of eigenvalues of finite multiplicities. In order to prove the second property, consider the operator $A(\lambda)=(\lambda-\lambda_0) B(\lambda)$. It is analytic in a neighborhood of $\lambda=\lambda_0$ and has exactly $p$ eigenvalues that  do not vanish at $\lambda_0$. Let $L_\lambda$ be the $p$-dimensional space spanned by the corresponding eigenfunctions of operator $A(\lambda)$. $L_\lambda$ is analytic in a neighborhood of $\lambda_0$ due to the above-mentioned  property of analytic self adjoint operators. By using $L_\lambda$ and its orthogonal complements, one can write the original operator $B(\lambda)$ in a neighborhood of $\lambda_0$ in a block form, where the block that  corresponds to $L_\lambda$ has a pole and the second block is analytic. After that, the statements of the second part of the lemma become obvious.

\qed

{\it Step 2. Relation between the set of the ITEs $\{\lambda_i^T\}$ and the eigenvalues $\mu_j=\mu_j(\lambda)$}.
Denote by $n^-(\lambda),~ \lambda\notin\{ \lambda_i\}\bigcup\{ \lambda^n_i\}\bigcup\{\lambda^T_i\}$,
the number of {\it negative} eigenvalues $\mu_j(\lambda)$ of the operator $B(\lambda)$. From (\ref{minf}) and Lemmas \ref{lemmaShubin} and \ref{hardDayLemma} it follows that this number is finite for each $\lambda$.


 Let us evaluate the difference $n^-(\lambda')-n^-(\alpha)$ by moving
 $\lambda$ from $\lambda=\alpha$ to $\lambda=\lambda'>\alpha$. Here $\alpha>0$ is the constant defined in (\ref{countdef}).
  The eigenvalues    $\mu_j(\lambda)$ are meromorphic functions of   $\lambda$,
 the number of negative eigenvalues $\mu_j(\lambda)<0$ changes only when some of them pass through
 the `edges' of the interval $\mathbb R^-_\mu=(-\infty,0)$. Denote by $n_1(\lambda')$
 the change in  $n^-(\lambda')-n^-(\alpha)$ due to the eigenvalues going through $\mu=-\infty$
 when $\lambda$ moves from $\alpha$ to
 $\lambda '$. Similarly, denote by $n_2(\lambda')$ the change in $n^-(\lambda')-n^-(\alpha)$ due to the eigenvalues going through $\mu=0$ when $\lambda$ moves from $\alpha$ to
 $\lambda '$. Then
\begin{equation}\label{n10}
n^-(\lambda')-n^-(\alpha)=n_1(\lambda')+n_2(\lambda').
\end{equation}

Note that the annihilation or the birth of $\mu_j(\lambda)$ at $\mu=-\infty$ may occur only when $\lambda$ passes through a pole $\lambda=\lambda_0$ of an eigenvalue $\mu_j(\lambda)$.
Let us denote by $\delta n_1(\lambda_0)$ the jump of $n_1$ at a pole $\lambda=\lambda_0$ of the operator $B(\lambda)$  due to some of  $\mu_j$ going through negative infinity.
\begin{lemma}\label{ms}
The following relation holds for every pole $\lambda=\lambda_0>0$ of the operator $B(\lambda)$:
\begin{equation}\label{s12}
\delta n_1(\lambda_0)=s^--s^+,
\end{equation}
where $s^-$ and $s^+$ are the numbers of negative and, respectively, positive eigenvalues of the residue $P_{\lambda_0}$ of operator $B(\lambda)$. When (\ref{28111}) holds, the latter relation becomes
\begin{equation}\label{mainlemma}
\delta n_1(\lambda_0)=\sigma(m_n-m_0),
\end{equation}
where $m_n$ and $m_0$ are ranks of the residues of the operators  $F_n(\lambda)$ and $F(\lambda)$, respectively, at the pole.
\end{lemma}
\noindent
{\bf Proof.} Recall that the eigenvalue $\mu_j(\lambda)$ may have poles only of the first order (see Lemma \ref{lemmasigma}). If an eigenvalue $\mu=\mu_j(\lambda)$ has a pole at $\lambda=\lambda_0>0$ with a positive residue,
then $\lim_{\lambda\to\lambda_0\pm 0}\mu_j(\lambda)=\pm\infty$, and therefore $\mu_j(\lambda)$ leaves
the negative semi-axis $\mu<0$ through $-\infty$ when $\lambda\to \lambda_0^-$.
Similarly, if the residue is negative, then the eigenvalues enter the semi-axis when $\lambda= \lambda_0^+$. This proves (\ref{s12}).

It is enough to prove (\ref{mainlemma}) when $F_n(\lambda)$ has a pole at  $\lambda=\lambda_0$. The case when  $F(\lambda)$ has a pole is similar.
Thus we assume that
\begin{equation}\label{ffn}
F_n(\lambda)=\frac{P_{\lambda_0}}{\lambda-\lambda_0} + F^0_n(\lambda),
\end{equation}
where $P_{\lambda_0}$ is a finite-dimensional (of rank $m_n$) operator and $F_n^0(\lambda)$ is a smooth operator in a neighborhood of $\lambda_0$. Operator $P_{\lambda_0}$ has exactly $m_n$ non-zero eigenvalues (their number coincides with the rank). We will show below that all of them are positive.
Then the residue $P=-\sigma DP_{\lambda_0}D$ of operator $B(\lambda)$ also
has exactly $m_n$ non-zero eigenvalues and their signs coincide with the sign of $-\sigma$. Hence (\ref{mainlemma}) follows from (\ref{s12}) and the last statement of Lemma \ref{hardDayLemma}. It remains to show that $P_{\lambda_0}\geq 0$.
The latter is an obvious consequence of (\ref{ffn}) and the following important statement \cite{fried}:  operator $\frac{d}{d\lambda}F_n(\lambda)$ is negative at every $\lambda$ that is not a pole of the operator. The proof of Lemma \ref{ms} is complete, but we will recall the proof of the statement from \cite{fried} to have all the details readily available to the reader.

Let $u=u(\lambda)$ be the solution of the equation $\Delta u + \lambda n(x)u=0$ in $\mathcal O \backslash \mathcal V$  with the Dirichlet data $\varphi$
at the boundary $\partial \mathcal O$ and let $u=0$ on $\partial \mathcal V$.
Its derivative $u'=\frac{d u(\lambda )}{d \lambda}$ satisfies the equation
$$
(\Delta  + \lambda n) u ' + n u = 0.
$$
Let us multiply this equation by $\overline{u}$, integrate over $\mathcal O \backslash \mathcal V$
and apply Green's formula. Since $u '=0$ on the boundary, we obtain that
$$
\int_{\partial \mathcal O} \frac{\partial u '}{\partial \nu} \overline{u}dS + \int_{\mathcal O \backslash \mathcal V}n(x) |u|^2 dx=0,
$$
which can be rewritten as
$$
\frac{d}{d\lambda} (F_n(\lambda) \varphi,\varphi)= -\int_{\mathcal O \backslash \mathcal V}n(x) |u|^2 dx<0.
$$
Thus, $\frac{d}{d\lambda}F_n(\lambda)<0$.

\qed

By summation of inequalities (\ref{mainlemma}) over all the poles $\lambda_0$ on the interval $(\alpha,\lambda)$, we obtain the following relation:
\begin{equation}\label{n1}
n_1(\lambda) = \sigma(N_n(\lambda)-N(\lambda)),\quad \lambda>\alpha,
\end{equation}
 where $N_n$ and $N$ are the counting functions defined in (\ref{countdef}). From (\ref{n10}) and (\ref{n1}) it follows that
\begin {equation}\label{nnn}
n^-(\lambda)-n^-(\alpha)+\sigma(N(\lambda)-N_n(\lambda)) = n_2(\lambda),\quad \lambda>\alpha .
\end{equation}

Due to Lemma \ref{lemma11}, the value of the counting function $N_T(\lambda)$ for the ITEs is equal to the number of zero values of all the eigenvalues $\mu_j(\tau),~j=1,2,... ~,$ when $\tau$ belongs to the interval $(\alpha,\lambda)$. Function $n_2(\lambda)$ also counts the number of zero values of $\mu_j(\tau),~j=1,2,... ~,$ when $\tau$ changes from $\alpha$ to $\lambda$, but $n_2(\lambda)$ counts these zeroes with coefficients $\pm 1$ or $0$. The choice of this coefficient depends on whether the corresponding $\mu_j(\tau)$ enters the semi-axis $\mathbb R_\mu^-$, exits it or does not change location with respect to the semi-axis when $\tau$ changes from $\alpha$ to $\lambda$ and passes through the point where $\mu_j(\tau)=0$. Thus  $N_T(\lambda)\geq n_2(\lambda)$.
 This and
 (\ref{nnn}) justify (\ref{mainIneq}) since $n^-(\lambda)\geq 0$.

\textit{Step 3. The case when (\ref{28111}) is violated.} We have additional ITEs in this case, and the following statement (which is also an immediate consequence of the definition of the ITEs) replaces Lemma \ref{lemma11}:
\begin{lemma}\label{ites}
A point $\lambda=\lambda_0$ is an ITE if and only if the operator $B(\lambda_0)$ has a non-trivial kernel or the following two conditions hold:

1) $\lambda=\lambda_0$ is an eigenvalue of the Dirichlet problem for $-\Delta$ and for equation (\ref{AnoneB}), i.e.,   $\lambda=\lambda_0$ is a pole for both $F(\lambda)$ and $F_n(\lambda)$.

2) The ranges of the residues of operators $F(\lambda)$ and $F_A(\lambda)$ at the pole $\lambda=\lambda_0$ have a non trivial intersection.

Moreover, the multiplicity of the interior transmission eigenvalue $\lambda=\lambda_0$ in all cases is equal to $m_1+m$, where $m_1$ is the dimension of the
kernel of the operator $B(\lambda_0)$, and   $m$ is the
dimension of the intersection of the ranges of the residues of operators $F(\lambda)$ and $F_n(\lambda)$  at the pole $\lambda=\lambda_0$ ($m=0$ if $\lambda=\lambda_0$ is not a pole).
\end{lemma}

Equality  (\ref{mainlemma}) must by replaced now by the following inequality, which is valid for every pole $\lambda=\lambda_0>0$ of operator $B(\lambda)$:
\begin{equation}\label{mainlemma2}
|\delta n_1(\lambda_0)-\sigma(m_n-m_0)|\leq m.
\end{equation}
Indeed, (\ref{s12}) remains valid in our case, but now we can not find $s^\pm$ explicitly. However, it is not difficult to show that $|(s^--s^+)-\sigma(m_n-m_0)|\leq m$, which leads to (\ref{mainlemma2}). The latter inequality follows easily from the fact (which can be found in the proof of Lemma \ref{ms}) that $-\sigma P \geq 0$ on the space $DV_n^\perp$ and $-\sigma P \leq 0$ on the space $DV_0^\perp$. Here $D$ is the operator defined in (\ref{bla}), $V_n$ and $V_0$ are
the ranges of the residues of $F_n$ and $F$, respectively, and $V_n^\perp,~V_0^\perp$ are the subspaces of the elements in $V_n,~V_0$ that are orthogonal to $V_n\bigcap V_0$. If more details are needed, they can be found in \cite[Lemma 2.4]{lakvain6}.

We will call  $\lambda=\lambda_0$ a singular ITE, and $m$ will be called its multiplicity, if the last two conditions of Lemma \ref{ites} hold. Let us denote by $R(\lambda)$ the counting function of the singular ITEs (the number, with multiplicities taken into account, of the singular ITEs whose values do not exceed $\lambda$).

By summation of inequalities (\ref{mainlemma2}) over all the poles $\lambda_0$ on the interval $(\alpha,\lambda)$, we obtain the following analogue of (\ref{n1}):
\begin{equation}\label{n12}
|n_1(\lambda)-\sigma(N_n(\lambda)-N(\lambda))|\leq R(\lambda),
\end{equation}
which leads to the following analogue of (\ref{nnn}):
\begin {equation}\label{nnn2}
n^-(\lambda)-n^-(\alpha)+\sigma(N(\lambda)-N_n(\lambda))\leq R(\lambda)+n_2(\lambda).
\end{equation}
It remains to note that $N_T(\lambda) \geq n_2(\lambda)$ when only non-singular ITEs are counted (the non-singular ITEs are related to the non-trivial kernels of $B(\lambda)$ and can be compared to $n_2(\lambda)$), and therefore, $N_T(\lambda) \geq n_2(\lambda)+R(\lambda)$ if all the ITEs are counted. This and (\ref{nnn2}) imply (\ref{mainIneq}).

\qed

\section{Calculation of the full symbol of the D-to-N operator.
}\label{mmm}
 Consider the problem
 \begin{equation}\label{sist}
 \left \{
 \begin{array}{l}
 Au=0,\quad x \in \mathcal O, \\
 u=\varphi, \quad x \in \partial \mathcal O,
 \end{array}
 \right .
 \end{equation}
 where $A$ is an elliptic differential operator of the second order with infinitely smooth coefficients, and the boundary $\partial \mathcal O$ is also infinitely smooth.

\begin{theorem} \label{ttt}
Let problem (\ref{sist}) be uniquely solvable. Then the operator $F:\varphi\to \frac{\partial u}{\partial \nu}$ is an elliptic pseudo-differential operator (p.d.o.) on $\partial\mathcal O$ of order one and its full symbol can be easily found by the procedure described below.
\end{theorem}

This theorem is a particular case of a more general statement proved in \cite{VG2}. An elliptic system $A$ of an arbitrary order is considered there with two different boundary operators $B_1$ and $B_2$ such that each of them complements $A$ to an elliptic boundary value problem. It is proved there that the operator $F:B_1u\to B_2u$ is an elliptic p.d.o. on $\partial \mathcal O$, and the full symbol of this operator is calculated. In particular, the results of \cite{VG2} imply that the full symbol of the D-to-N operator $F$ has the following form (these calculations can be also found in the later publications  \cite[Ch.VII]{Eskin},\cite{Lee}).

Let $V$ be a small neighborhood of a point on $\partial \mathcal O$ with local coordinates $(y_1,\ldots,y_{d-1},t)$  such that $\partial \mathcal O$ in $V$ is given by the equation $t=0$, and $V\bigcap\mathcal O$ is defined by $ |y|^2+t^2<\varepsilon^2, ~t>0$. Let $A(y,t,i\frac{\partial}{\partial y},i\frac{\partial}{\partial t})$ be the operator $A$ rewritten in local coordinates $(y,t)$. The symbol of this operator in new coordinates is $A(y,t,\xi, \tau)$. A function $\Phi(z,t,\xi,\tau)$ will be called generalized homogeneous of order $\gamma$ if
$$
\Phi(\kappa z,\kappa t,\kappa \xi, \kappa \tau) = \kappa^{\gamma}\Phi(z,t,\xi,\tau)
$$
for every $\kappa>0$. For each $x\in V\bigcap \partial \mathcal O$ and $N>0$, the symbol of $A$ can be written in the form
\begin{equation}\label{T}
A(y,t,\xi,\tau) = \sum_{j=0}^N A_j(x,y-y(x),t,\xi,\tau)+
A_N'(x,y-y(x),t,\xi,\tau),
\end{equation}
where $A_j(x,z,t,\xi,\tau)$ are generalized homogeneous polynomials in $(z,t,\xi,\tau)$ of order $2-j$ and
\[
|A_N'(x,\kappa z,\kappa t,\kappa \xi, \kappa \tau)|\leq C(x,z,t,\xi, \tau)\kappa^{1-N} \quad \text{when}\quad |z|^2+t^2<1 ,~~\kappa\to 0.
\]
In order to obtain expansion (\ref{T}), one needs to write $A$ as a Taylor series in $z$ and $t$ centered at the point $(y(x),0)$ and group together the terms of the same order.

Let
\begin{eqnarray}\label{lapl}
\widetilde{A}_j = A_j(x,-i\frac{\partial}{\partial \xi},t,\xi,i\frac{\partial}{\partial t}).
\end{eqnarray}
Consider the following recursive  system of ODEs on the half line $t>0$ (which depend on the parameters $x$ and $\xi$):
\begin{eqnarray}
\widetilde{A}_0 E_0 (x,t,\xi) =0,\label{appsist1}\\
 \widetilde{A}_0 E_1 (x,t,\xi) = -\widetilde{A}_1 E_0 (x,t,\xi)\label{appsist2}\\
 \widetilde{A}_0 E_2 (x,t,\xi)= -\widetilde{A}_1 E_1 (x,t,\xi) - \widetilde{A}_2 E_0 (x,t,\xi) \label{appsist3}\\
 \ldots \nonumber \\
 \widetilde{A}_0 E_j (x,t,\xi)= -\widetilde{A}_1 E_{j-1} (x,t,\xi) -\ldots- \widetilde{A}_j E_0 (x,t,\xi)\label{appsist4}.
\end{eqnarray}
From the ellipticity of operator $A$ it follows that this system has a unique solution $\{E_i\},i=1,2,...,$ in the class of functions that decay at infinity and satisfy the following initial data:
$$
E_0(x,t,\xi)|_{t=0}=1, \quad E_j(x,t,\xi)|_{t=0}=0,~~j>0.
$$
 The full symbol $F(x,\xi)$ of $F$  is given by the following asymptotic series \cite[Th.14]{VG2}:
\begin{equation}\label{ps}
F(x,\xi)=-\sum_{j=0}^\infty( \frac{d}{d t} E_{j})|_{t=0} .
\end{equation}

\noindent
{\bf Example.} As an example, consider $A=-\Delta+n(x)\lambda$. We introduce local coordinates $(y,t)$ where $y=(y_1,...y_{d-1}),~y=y(x),$ are local coordinates on $\partial \mathcal O$ and $t$ is the distance between a point $x$ and $\partial \mathcal O$. Then the principal symbol of the operator $A$ when $t=0$ is equal to
 $$
 A_0=A_0(x,0,0,\xi,\tau)=\tau ^2+\sum g^{i,j}(y)\xi_i\xi_j,
 $$
 where $\sum g_{i,j}(y)dy_idy_j$ is the first fundamental form (the first quadratic form) on $\partial \mathcal O$.
We will call
\begin{equation}\label{xis}
|\xi^* |=(\sum g^{i,j}(y)\xi_i\xi_j)^{1/2}
\end{equation}
the length of the co-vector $\xi$. It depends on $x$ and the choice of the local coordinates. Then  $E_0=e^{-t|\xi^* |}$, and the principal symbol of the D-to-N operator $F$ is $| \xi^* |$ where $(x,\xi)$ belongs to the co-tangent bundle $T^*(\partial \mathcal  O)$.

\section{ Proofs of main Lemmas.}

\noindent
\textbf{ Proof of Lemma \ref{lemmasigma}.} It is easy to use the results of the previous section and obtain formulas (\ref{Grcond3}), (\ref{grcond3a}) for the principal symbol of $F_n-F$. It will take much longer to justify the analytic properties of this operator in specific spaces indicated in (\ref{diff}).

\textit{Step 1. Proof of the first part of the lemma when $\lambda$ belongs to a disk $|\lambda-\lambda_0|<a$ that is free of eigenvalues of both Dirichlet problems: for equation (\ref{Anone0}) and for equation (\ref{AnoneB})}. Theorem \ref{ttt} and calculations of the symbol from the previous section can be applied in this case.

If $A=-\Delta+n(x)\lambda$, we denote operators $\widetilde{A}_k$ and functions $E_k$ introduced in the previous section by $\widetilde{A}^n_k$ and $E_k^n$, respectively, and we preserve the previous notations (without index $n$) if $n\equiv 1$.
Obviously,
$\widetilde{A}_0^n=\widetilde{A}_0, ~\widetilde{A}_1^n=\widetilde{A}_1$, and $\widetilde{A}_2^n-\widetilde{A}_2=\lambda (1-n(x))$. For the sake of transparency of the proof, we will assume that $\widetilde{A}_0=\widetilde{A}^n_0=-\frac{d^2}{d t^2}+|\xi^*|^2$, see the example above (all the calculations could be easily made in arbitrary local coordinates). Hence $E_0^n=E_0=e^{-t|\xi^*|},~E_1^n=E_1$, and
\begin{equation}\label{eq1}
[-\frac{d^2}{d t^2}+|\xi^*|^2](E_2^n-E_2)=\lambda (n-1) e^{-t|\xi^*|}.
\end{equation}
Function  $$
Y(t)=\frac{\lambda (1-n(x))}{2|\xi^*|^2}te^{-t|\xi^*|}.
$$
is a particular solution of equation (\ref{eq1}). It vanishes at $t=0$ and at infinity, i.e., $E_2^n-E_2=Y(t)$. Thus the first two terms of the full symbol of the operator  $F_n-F$ are zeroes, and the next one is equal to $-\frac{d}{dt}Y(t)|_{t=0}$. The latter expression coincides with (\ref{Grcond3}). Hence, $F_n-F$ is a p.d.o. of the order $-1$ with the principal symbol (\ref{Grcond3}).

Let now $n(x)= 1,~\frac{\partial n}{\partial \nu}\neq 0$ at $\partial\mathcal O$. In this case,  $\widetilde{A}_j^n=\widetilde{A}_j,~j=0,1,2,$ and  $\widetilde{A}_3^n-\widetilde{A}_3=t\lambda \frac{\partial n}{\partial \nu}$ (note that $t$ and $\nu$ have different directions). Hence, $E_j^n=E_j,~j=0,1,2$, and the following equation holds for $E_3^n-E_3$:
\begin{equation}\label{eq1a}
[-\frac{d^2}{d t^2}+|\xi^*|^2](E_3^n-E_3)=-t\lambda \frac{\partial n}{\partial \nu} e^{-t|\xi^*|}.
\end{equation}

The solution of (\ref{eq1a}) that decays at infinity and vanishes at $t=0$  has the form
\begin{equation}\label{eq1b}
E_3^n-E_3=-\lambda \frac{\partial n}{\partial \nu}(\frac{t^2}{4|\xi^*|}+\frac{t}{4|\xi^*|^2})e^{-t|\xi^*|}.
\end{equation}
 Hence the first non-zero term of the full symbol of the operator $F_n-F$ is given by (\ref{grcond3a}).

We proved (\ref{diff})-(\ref{grcond3a}) for $\lambda$ in a disk $|\lambda-\lambda_0|<a$ that does not contain poles of $F_n$ and $F$. Now we are going to study the analytic properties of operator (\ref{diff}). We will do it when $n(x)= 1,~\frac{\partial n}{\partial \nu}\neq 0$ at $\partial\mathcal O$. The case of $n(x)\neq 1,~x\in\partial\mathcal O,$ can be studied similarly.
One of the difficulties in the proof of the analyticity of operator (\ref{diff}) is related to the fact that the range of operators $F_n$ and $F$ (i.e. space $H^{1/2}(\partial\mathcal O))$ is wider than the range of the difference of these operators indicated in (\ref{diff}). In order to prove the analyticity of operator (\ref{diff}) we will use the Taylor expansion of $F_n-F$ at $\lambda=0$ since the coefficients $\lambda^j,~j>0,$ in the Taylor expansion have better smoothing properties than the operator itself. Indeed, let us write the solution of the problem (\ref{AnoneB}) with the Dirichlet condition $v|_{\partial\mathcal O}=\phi$ in the form
\begin{equation}\label{v0}
 v=v_0(x)+\lambda v_1(x)+ w(\lambda,x),
\end{equation}
 where $v_0, v_1$ do not depend on $\lambda$ and
\begin{equation}\label{v01}
 -\Delta v_0=0,~x\in \mathcal O\backslash \mathcal V,~v_0|_{\partial\mathcal V}=0,~v_0|_{\partial\mathcal O}=\phi;
 \end{equation}
\begin{equation}\label{v02}
 -\Delta v_1=n(x)v_0,~x\in \mathcal O\backslash \mathcal V,~~v_1|_{\partial\mathcal V}=v_1|_{\partial\mathcal O}=0;
\end{equation}
\begin{equation}\label{v03}
 -\Delta w-\lambda n(x)w=\lambda^2 n(x)v_1,~x\in \mathcal O\backslash \mathcal V,~w|_{\partial\mathcal V}=w|_{\partial\mathcal O}=0.
\end{equation}
 Then
 \[
\| v_0\|_{H^2}\leq C \|\phi\|_{H^{3/2}(\partial\mathcal O)},~~\| v_1\|_{H^4}\leq C \| v_0\|_{H^2}\leq C\|\phi\|_{H^{3/2}(\partial\mathcal O)},~~\| w\|_{H^6}\leq C \|\phi\|_{H^{3/2}(\partial\mathcal O)},
 \]
 and the operator $\Gamma:H^{3/2}(\partial\mathcal O)\to H^6(\mathcal O\backslash \mathcal V)$ that maps $\phi$ into $w$ is bounded and analytic in $\lambda$ when $|\lambda-\lambda_0|<a$. By taking the normal derivative on
$\partial\mathcal O$ in both sides of (\ref{v0}), we obtain the following representation for $F_n(\lambda)$:
\[
F_n(\lambda)=F_n^0+\lambda F_n^1+ F_n^2(\lambda),~~~F_n^2(\lambda):H^{3/2}(\partial\mathcal O)\to H^{9/2}(\partial\mathcal O), \quad |\lambda-\lambda_0|<a,
\]
where operators $F_n^0$ and $ F_n^1$ do not depend on $\lambda$, and $ F_n^2$ depends on $\lambda$ analytically. A similar representation is valid for $F(\lambda)$. Hence
\begin{equation}\label{analyt}
F_n(\lambda)-F(\lambda)=G_n^0+\lambda G_n^1+G_n^2(\lambda),~~~G_n^2(\lambda):H^{3/2}(\partial\mathcal O)\to H^{9/2}(\partial\mathcal O), \quad |\lambda-\lambda_0|<a,
\end{equation}
where the operators $G_n^0$ and $ G_n^1$ do not depend on $\lambda$, and $ G_n^2$ is analytic in $\lambda$.

 Since the operator $F_n-F:H^{3/2}(\partial\mathcal O)\to H^{7/2}(\partial\mathcal O)$ is bounded for each $\lambda$ (it is  a p.d.o. of order $-2$), from (\ref{analyt}) it follows that operators $G_n^0,~G_n^1$ are bounded in the same spaces. Thus  (\ref{analyt}) implies that operator (\ref{diff}) is analytic when $|\lambda-\lambda_0|<a$.

\textit{Step 2. Proof of the first part of lemma when $\lambda_0$ is a pole of either $F_n$ or $F$ (or both operators).} We could repeat the previous arguments, but we will need to apply the resolvent $ (-\Delta -\lambda n(x))^{-1}$ to both sides of (\ref{v03}) in order to obtain $w$. Thus operator $G_n^2$ will have a pole of the first order at $\lambda=\lambda_0$, whose residue can be expressed through the residue of the resolvent.
This approach leads to a slightly weaker result than the one stated in the lemma: it gives the description of the residue, but the regular part of $F_n-F$ will be represented as a p.d.o. plus a smoother operator. While this result is sufficient for all applications in this paper, we decided to spend a little more time and prove the exact statement of Lemma \ref{lemmasigma}, i.e., to prove that the regular part is a p.d.o. up to an infinitely smoothing operator, and the principal symbol is given by (\ref{Grcond3}), (\ref{grcond3a}).

In order to study the case when $\lambda_0$ is a pole, we perturb equations (\ref{Anone0}), (\ref{AnoneB}) by adding an infinitely smooth term $q(x)\in C^\infty$ to the potential $n(x)$ in these equations. For example, equation (\ref{Anone0}) will now take the form $-\Delta u-\lambda (1+q(x))u=0.$ The goal of these perturbations is to get rid of the eigenvalue at $\lambda=\lambda_0$, while preserving the symbols of the regular parts of operators $F_n$ and $F$. We may need to choose different potentials for these two equations, but we will use the same notation $q$ for both equations (we never compare potential terms below).

We choose terms $q$ that satisfy the following two requirements: $q$ vanish in a neighborhood of $\partial\mathcal O$, and the homogeneous Dirichlet problems for both equations (\ref{Anone0}) and (\ref{AnoneB}) with the terms $q$ added have only trivial solutions when $\lambda=\lambda_ 0$. For example, in order to achieve the second requirement in the case of real valued $n(x)$, one can take $q(x)=iq_1(x),~q_1(x)\geq 0.$ If $n(x)$ is complex valued, one can take $q(x)=-\alpha(x)n(x)$, where $\alpha\in C^\infty, ~0\leq\alpha\leq 1,~ \alpha=1$ outside of a small enough neighborhood of $\partial\mathcal O, ~\alpha$ vanishes in a smaller neighborhood of $\partial\mathcal O$.

Since $\lambda=\lambda_0 $ is not an eigenvalue for $F_n$ or $F$, there exists an $a>0$ such that the disk $|\lambda-\lambda_ 0|<a$ does not contain eigenvalues of the Dirichlet problem for equations (\ref{Anone0}) or (\ref{AnoneB}) with the additional term $q$ added. Thus
the D-to-N operators $F_{n,q},~F_q, ~|\lambda-\lambda_ 0|<a,$ are defined for these equations.

Operators $F_{n,q},~F_q,~|\lambda-\lambda_ 0|<a,$ can be studied absolutely similarly to the operators with $q=0$. They are p.d.o. of the first order. Their full symbols can be constructed exactly as for operators with $q=0$. Moreover, since $q=0$ in a neighborhood of $\partial\mathcal O$, the construction of the full symbol of these operators does not depend on $q$ at all. One can also repeat the arguments leading to (\ref{analyt}) and obtain a similar representation for $F_{n,q}-F_q$ when
$\lambda\in \mathbb R^+$. We will need only to add the term  $q$ to $n$ in equations (\ref{v01})-(\ref{v03}). Hence
\begin{equation}\label{diff1}
F_{n,q}(\lambda)-F_q(\lambda):H^{\frac{3}{2}}(\partial\mathcal O)\to H^{\frac{3}{2}+s}(\partial\mathcal O),~~ |\lambda-\lambda_ 0|<a,
\end{equation}
 is an elliptic pseudo-differential operator of order $-s$, which is analytic in $\lambda$, and its principal symbol is given by (\ref{Grcond3}) or (\ref{grcond3a}).

 Note that operator (\ref{diff1}) differs from (\ref{diff}) by $I_1+I_2$, where $I_1=F_{n,q}(\lambda)-F_n(\lambda),~ I_2=F_{q}(\lambda)-F(\lambda)$. We are going to show that
\begin{equation}\label{diff11}
 I_1=F_{n,q}(\lambda)-F_n(\lambda)=\frac{P_{\lambda_0,n}}{\lambda-\lambda_0}+Q_n(\lambda),~~|\lambda-\lambda_ 0|<a,
 \end{equation}
 where integral kernels $P_{\lambda_0,n}(x,y),~Q_n(\lambda,x,y),~x,y\in \partial\mathcal O,$ of operators $P_{\lambda_0,n}$, $Q_n(\lambda)$ are infinitely smooth functions of their arguments, $P_{\lambda_0,n}$ does not depend on $\lambda$, $Q(\lambda,x,y)$ depends on $\lambda$ analytically, and $P_{\lambda_0,n}$ is a projection on the space spanned by the normal derivatives of the solutions of the homogeneous Dirichlet problem for equation (\ref{AnoneB}). Thus $I_1$ has a pole, but its principal part $Q_n$ is an infinitely smoothing operator. The validity of (\ref{diff11}) for arbitrary $n$ implies, in particular, its validity for $I_2$, where $n\equiv 1$. Hence (\ref{diff1}) and (\ref{diff11}) together justify the first statement of Lemma \ref{lemmasigma}. It remains only to prove (\ref{diff11}).

We have
$$ F_{n}(\lambda)\phi=\frac{\partial}{\partial \nu}v|_{\partial\mathcal O},~~F_{n,q}(\lambda)\phi=\frac{\partial}{\partial \nu}v_1|_{\partial\mathcal O},
$$
where $v,v_1$ are the solutions of the following problems (for shortness, we will assume that $\mathcal V=\emptyset$, but one can easily add $\mathcal V$ below)
\[
-\Delta v-\lambda n(x) v=0, ~~x\in \mathcal O,~~v|_{\partial\mathcal O}=\phi,
\]
\begin{equation}\label{v11}
-\Delta v_1-\lambda (n(x)+q(x)) v_1=0, ~~x\in \mathcal O,~~v_1|_{\partial\mathcal O}=\phi.
 \end{equation}
Thus, $(F_{n,q}- F_{n})\phi=\frac{\partial}{\partial \nu}w|_{\partial\mathcal O}$, where $w$ satisfies
\begin{equation}\label{w11}
-\Delta w-\lambda n(x)w=\lambda q(x)v_1, ~~x\in \mathcal O,~~w|_{\partial\mathcal O}=0.
 \end{equation}

 Hence
\[
(F_{n,q}- F_{n})\phi=\frac{\partial}{\partial \nu}\int_{\mathcal O}R_\lambda(x,y)\lambda qv_1(y)dy|_{\partial\mathcal O},
\]
where $R_\lambda(x,y)$ is the kernel of the resolvent $(-\Delta w-\lambda n(x))^{-1}$ (for the operator with zero Dirichlet condition). The resolvent has a pole of the first order at $\lambda=\lambda_0$.

Since problem (\ref{v11}) does not have eigenvalues in the disk $|\lambda-\lambda_ 0|<a$ and $q=0$ in a neighborhood of $\partial\mathcal O$, it follows that
$$
q(x)v_1=\int_{\partial\mathcal O}K(\lambda,x,y)\phi(y)ds_y,
$$ where the kernel $K$ is infinitely smooth and analytic in $\lambda$. The latter two formulas imply (\ref{diff11}) since in both formulas $x$ and $y$ are separated. The first statement of Lemma \ref{lemmasigma} is proved.

\textit{Step 3. Proof of the second part of the lemma.} Let $\mathcal V=\emptyset$. We note that (\ref{analyt}) remains valid when $\lambda_0=0$. In particular, $G_n^2(\lambda)$ is analytic at $\lambda=0$ in this case. Further, $G_n^0=0$ when $\mathcal V=\emptyset$ since functions $v_0$ determined by (\ref{v01}) are the same for both operators $F_n$ and $F$. Thus  (\ref{analyt}) takes the form
\begin{equation}\label{47}
F_n(\lambda)-F(\lambda)=\lambda G_n^1+G_n^2(\lambda),~~~G_n^2(\lambda):H^{3/2}(\partial\mathcal O)\to H^{9/2}(\partial\mathcal O), \quad |\lambda|<a\ll 1,
\end{equation}
where operator $ G_n^1$ does not depend on $\lambda$, and $ G_n^2$ is analytic in $\lambda$.

It is an obvious consequence of the definition of operators $F_n,~F$ that the operators $F_n(0)$ and $F(0)$ coincide. Thus $ G_n^2(0)=0$, and (\ref{47}) implies that the operator $G(\lambda)=\frac{F_n(\lambda)-F(\lambda)}{\lambda}$ can be extended analytically at $\lambda=0$. It remains only to show that $G(0)$ is Fredholm.

It is proved in the first part of the lemma that the operator $F_n(\lambda)-F(\lambda)$ is an elliptic p.d.o. of order $-s$ when $\lambda\neq 0 $ and is not a pole. Thus the operator
\begin{equation}\label{48}
\lambda G_n^1+G_n^2(\lambda):H^{3/2}(\partial\mathcal O)\to H^{3/2+s}(\partial\mathcal O), ~~0<|\lambda|<a,
\end{equation}
is Fredholm. Since $s=1$ or $2$, from (\ref{47}) and the Sobolev imbedding theorem it follows that the second term in the left-hand side of  (\ref{48}) is a compact operator (by the same reason, $(G^2)':=\lim_{\lambda\to 0}\lambda^{-1}G_n^2(\lambda)$ is compact). Thus $G_n^1$ is Fredholm. Since $(G^2)'$ is Fredholm, the operator $G(0)=G_n^1+(G^2)'$ is Fredholm.

\qed

 The proof of Lemma \ref{paramellTh} is based on the
parameter-ellipticity of equations (\ref{Anone0}) and (\ref{Anone}), which leads to the parameter-ellipticity of the operators $F_n$ and $F$. The main terms of their symbols are canceled when the difference $F_n-F$ is taken, but the lower order terms still inherit some properties of parameter-ellipticity. So, the proof of Lemma  \ref{paramellTh} relies essentially on evaluating the first few terms of the parameter-elliptic operators $F_n$ and $F$.

Recall that parameter-ellipticity means that the operator of multiplication by $\sqrt\lambda$ has the same order as  differentiation. More precisely, the full symbol of a parameter-elliptic p.d.o. is an asymptotic series of terms, which are homogeneous in $(|\xi^*|, k)$, where  $k=\sqrt{|\lambda|}$, with the main term (principal symbol) not vanishing when $|\xi^*|^2+k^2\neq 0$. Thus we will calculate the first three terms $f_j,~j=0,1,2,$ of the full symbol $\sum_{j=0}^\infty f_j$
of the parameter-elliptic p.d.o. $F_n$ (the terms $f_j=f_j(x,\xi,\lambda)$ are homogeneous functions of $\xi$ and $k$ of order $1-j$) before the proof of Lemma \ref{paramellTh}. We will assume that the transformation matrix from global to local coordinates is orthogonal on the boundary $\partial \mathcal O$ (this can always be done in dimensions $d\leq 3$). We will need the following result.

\begin{lemma}\label{symb} Let $\lambda\in\Lambda$, where $\Lambda$ is a closed sector of the complex plane that does not contain points of the set $\mathcal N$ defined in (\ref{en}).  Then the principal symbol $f_0$ of the parameter-elliptic p.d.o. $F_n$ has the form
\begin{equation}\label{prs}
f_0=\sqrt{|\xi^*|^2-n(x)\lambda}.
\end{equation}
If $n(x)\neq 1$ on $\partial \mathcal O$, then $f_1(x,\xi,0)$ does not depend on $n(x)$. If $n(x)\equiv 1$ on $\partial \mathcal O$, then
\begin{equation}\label{f1}
f_1=g_1(x,\xi,\lambda)+\frac{\frac{\partial n}{\partial \nu}\lambda}{4(|\xi^*|^2-\lambda)},
\end{equation}
where $g_1(x,\xi,\lambda)$ does not depend on $n(x)$. The term $f_2$ is $n$-independent when $\lambda=0.$
\end{lemma}
\noindent
\textbf{Remark.} Obviously, $f_0\neq 0$ when $\lambda\in\Lambda$ and $|\xi^*|^2+{|\lambda|}\neq 0$, i.e., $F_n$ is parameter-elliptic.

\noindent
\textbf{Proof.} The full symbol $\sum f_j$ can be found absolutely similarly to calculations (\ref{T})-(\ref{ps}) of the parameter independent symbol of $F_n$. However, now all the expansions in generalized homogeneous terms must include $\sqrt\lambda$ with the same weight as the weight of $|\xi^*|$. We will use notations $\widetilde{A}_{j,\lambda}$ and $E_{j,\lambda}$ for operators $\widetilde{A}_j$ and functions $E_j$, respectively, in order to stress that they are different now. In particular,
\[
\widetilde{A}_{0,\lambda}=-\frac{d^2}{d t^2}+|\xi^*|^2-n(x)\lambda.
\]
From (\ref{appsist1}) it follows that, for $n(x)>0$,
\[
E_{0,\lambda}=e^{-t\sqrt{|\xi^*|^2-n(x)\lambda}}, \quad \lambda \in \Lambda, \quad \text{Re}\sqrt{|\xi^*|^2-n(x)\lambda}>0,~~\text{when}~~|\xi^*|^2+|\lambda|\neq 0.
\]
Thus the principal symbol $f_0=-\frac{\partial}{\partial t}E_{0,\lambda}|_{t=0}$ of the operator $F_n$ is equal to (\ref{prs}).

 Similarly, one can evaluate $\widetilde{A}_{1,\lambda}$. It consists of two parts: the contribution from the Laplacian (which is defined in (\ref{lapl})) and the contribution from $-\lambda n(x)$. Namely,
\[
\widetilde{A}_{1,\lambda}=\widetilde{A}_1-\lambda \widetilde{B}_1,\quad \widetilde{B}_1=-\frac{\partial n}{\partial \nu}t+\langle -i\frac{\partial}{\partial\xi},\nabla_y n\rangle.
\]

Our next step is to evaluate
$E_{1,\lambda}$ and the second term $f_1=f_1(x,\xi,\lambda)=-\frac{d}{d t}E_{1,\lambda}|_{t=0}$ of the full symbol of the parameter-elliptic operator $F_n$. Equation (\ref{appsist2}) for $E_{1,\lambda}$ has the form $\widetilde{A}_{0,\lambda}E_{1,\lambda}=-\widetilde{A}_{1,\lambda}E_{0,\lambda}$. The right-hand side here and the operator  $\widetilde{A}_{0,\lambda}$ do not depend on $n(x)$ when $\lambda=0$. Thus $E_{1,0}$ and $f_1(x,\xi,0)$ are independent of $n(x)$.

Assume now that $n(x)\equiv 1$ on $\partial \mathcal O$. Then $\nabla_y n=0$, and $E_{1,\lambda}$ can be written as $E_{1,\lambda}=G+H$, where $G$ and $H$ are the solutions of the equations:
\[
(-\frac{d^2}{d t^2}+|\xi^*|^2-\lambda)G=-\widetilde{A}_{1}e^{-t\sqrt{|\xi^*|^2-\lambda}}, \quad (-\frac{d^2}{d t^2}+|\xi^*|^2-\lambda)H=-\lambda\frac{\partial n}{\partial \nu}te^{-t\sqrt{|\xi^*|^2-\lambda}}
\]
that vanish at $t=0$ and when $t\to \infty$. Obviously, $G$ and $g_1=-\frac{d}{d t}G|_{t=0}$ do not depend on $n(x)$. Solving equation for $H$, we obtain
\[
H=-\frac{\lambda }{4}(\frac{t^2}{\sqrt{|\xi^*|^2-\lambda}}+\frac{t}{|\xi^*|^2-\lambda})\frac{\partial n}{\partial \nu} e^{-t\sqrt{|\xi^*|^2-\lambda}}~,
\]
and therefore,
\[
-\frac{d}{d t}H|_{t=0}=\frac{\frac{\partial n}{\partial \nu}\lambda}{4(|\xi^*|^2-\lambda)}.
\]
Thus (\ref{f1}) is proved. In order to find $f_2$, we need to solve the equation $\widetilde{A}_{0,\lambda}E_{2,\lambda}=-\widetilde{A}_{1,\lambda}E_{1,\lambda}-\widetilde{A}_{2,\lambda}E_{0,\lambda}$. One can easily check that the right-hand side here and  the operator  $\widetilde{A}_{0,\lambda}$ do not depend on $n(x)$ when $\lambda=0$. Thus $E_{2,0}$ and $f_2(x,\xi,0)=-\frac{d}{d t}E_{2,0}|_{t=0}$ are independent of $n(x)$.

\qed

\noindent
\textbf{Proof of lemma \ref{paramellTh}.} The following fact will be used below. Let  $\Phi$ be a parameter-elliptic p.d.o. on $\partial \mathcal O$ of order $s$ when $\lambda$ belongs to a  closed sector $\Lambda$ in the complex $\lambda$-plane. Then, \cite[th.4.4.6]{A1},\cite{Grubb}
for each $m\in \mathbb R$, the operator
\begin{equation}
\Phi:H^{m,k}(\partial \mathcal O)\to H^{m+s,k}(\partial \mathcal O),\quad  \lambda\in\Lambda'=\Lambda\bigcap\{|\lambda|>1\}, \quad k=\sqrt{|\lambda|},
\end{equation}
is bounded, as well as its inverse for sufficiently large $|\lambda|$, and their norms can be estimated uniformly in $\lambda$. Here $H^{m,k}(\partial \mathcal O)$ is the Hilbert space with the norm defined in (\ref{norm}).

Denote by $K_n(\lambda)$ the p.d.o. on $\partial \mathcal O$, which is defined by the following two properties. Its construction uses the same partition of unity on  $\partial \mathcal O$ that was used to define $F_n(\lambda)$. The symbol of  $K_n(\lambda)$ in each local chart $V\in \partial \mathcal O$ consists of the first two terms of the full symbol of the parameter-elliptic operator $F_n(\lambda)$. Then the operator  $K_n(\lambda)$ is also parameter-elliptic of order one. The difference $F_n(\lambda)-K_n(\lambda)$ has order $-1$, i.e.,
\begin{equation}\label{nb2}
\parallel (F_n(\lambda)-K_n(\lambda))f\parallel_{H^{m+1,k}}\leq C\parallel f\parallel_{H^{m,k}}, \quad \lambda\in\Lambda', \quad k=\sqrt{|\lambda|}.
\end{equation}

Denote by $K(\lambda)$ the operator $K_n(\lambda)$ with $n\equiv 1$. Lemma \ref{symb} implies that the symbol $p(x,\xi,\lambda)$ of the operator $K_n(\lambda)-K(\lambda)$ in $V$ is given by the expression
\begin{eqnarray}
p(x,\xi,\lambda)=\sqrt{|\xi^*|^2-n(x)\lambda}-\sqrt{|\xi^*|^2-\lambda}+f_1(x,\xi,\lambda;n)-f_1(x,\xi,\lambda;1)\nonumber \\
=\frac{(1-n(x))\lambda}{\sqrt{|\xi^*|^2-\lambda}+\sqrt{|\xi^*|^2-n(x)\lambda}}+[f_1(x,\xi,\lambda;n)-f_1(x,\xi,\lambda;1)].
\label{s22}
\end{eqnarray}
Here $f_1$ is the function defined in Lemma \ref{symb}, but we added one more argument in the notation of this function in order to stress that $f_1$ depends on $n=n(x)$.

Assume now that $n(x)\neq 1$ on $\partial \mathcal O$. Function (\ref{s22}) is the sum of two terms, which are generalized homogeneous functions of order one and zero, respectively. It is important that function $p$ can be written in the form
\begin{equation}\label{kkk}
p(x,\xi,\lambda)=\lambda[p_1(x,\xi,\lambda)+p_2(x,\xi,\lambda)],\quad \lambda\in\Lambda',
\end{equation}
where the functions $p_1$ and $p_2$ are smooth when $|\xi^*|^2+|\lambda|\neq 0$, generalized homogeneous of order $-1$ and $-2$, respectively, and $p_1\neq 0$ when $|\xi^*|^2+|\lambda|\neq 0$. Indeed, the properties of $p_1$ are obvious, since the relation (which is obtained by equating the denominator in (\ref{s22}) to zero)
 \begin{equation}\label{k1k}
\sqrt{|\xi^*|^2-\lambda}=-\sqrt{|\xi^*|^2-n(x)\lambda}
\end{equation}
implies that $\lambda(1-n(x))=0$, i.e., $\lambda=0$. Then (\ref{k1k}) requires $\xi=0$.
Furthermore,
\[
f_1(x,\xi,\lambda;n)-f_1(x,\xi,\lambda;1)
\]
vanishes when $\lambda=0$ since Lemma \ref{symb} implies that $f_1(x,\xi,0;n)$ does not depend on $n$. Thus
\[
p_2=\lambda^{-1}[f_1(x,\xi,\lambda;n)-f_1(x,\xi,\lambda;1)], \quad |\xi^*|^2+|\lambda|\neq 0,
\]
is smooth. Hence (\ref{kkk}) holds.

From (\ref{kkk}) it follows that
\[
K_n(\lambda)-K(\lambda)=\lambda R(\lambda),
\]
where $R$ is a parameter-elliptic p.d.o. of order $-1$ when $\lambda\in\Lambda'$ (its symbol equals $\lambda^{-1}p=p_1+p_2$). Thus, the operator
\[
|\lambda|^{-1}[K_n(\lambda)-K(\lambda)]:H^{m,k}(\partial \mathcal O)\to H^{m+1,k}(\partial \mathcal O),\quad  \lambda\in\Lambda', \quad k=\sqrt{|\lambda|},
\]
is uniformly bounded in $\lambda$. This and (\ref{nb2}) imply (\ref{s1}). Let us prove (\ref{s2}). We have
 \begin{equation}\label{end}
F_n-F=K_n-K+(F_n-K_n)-(F-K)=R(\lambda)\{\lambda I+R^{-1}(\lambda)T_1\},\quad  \lambda\in\Lambda',\quad |\lambda|\gg 1,
\end{equation}
where $I$ is the identity operator and the operator $T_1=(F_n-K_n)-(F-K)$ has order~$-1$. It remains to note that
\[
\parallel R^{-1}(\lambda)f\parallel_{H^{m,k}}\leq C\parallel f\parallel_{H^{m+1,k}}, \quad \lambda\in\Lambda', \quad k=\sqrt{|\lambda|}\to\infty,
\]
i.e., $R^{-1}(\lambda)T_1$ is a bounded operator in $H^{m,k}$, and therefore (\ref{end}) leads to (\ref{s2}).

Assume now that $n(x)\equiv 1, \frac{\partial n}{\partial \nu}\neq 1$ on $\partial \mathcal O$. Then we denote by $K_n$ the p.d.o., whose symbol in each chart $V\in \partial \mathcal O$ consists of the first three terms of the full symbol of the parameter-elliptic operator $F_n(\lambda)$.
Then the operator $T_2=(F_n-K_n)-(F-K)$ has order~$-2$ and the operator $K_n-K$ has the form $K_n-K=\lambda R(\lambda)$, where $R(\lambda)$ is a parameter-elliptic p.d.o. of order $-2$ (its principal symbol is $\frac{\frac{\partial n}{\partial \nu}}{4(|\xi^*|^2-\lambda)}$, and the remaining part of the symbol is $\lambda^{-1}[f_2(x,\xi,\lambda;n)-f_2(x,\xi,\lambda;1)]$). Now, the second part of the statement of the theorem   follows easily from the representation:
\[
F_n-F=K_n-K+(F_n-K_n)-(F-K)=R(\lambda)\{\lambda I+R^{-1}(\lambda)T_2\},\quad  \lambda\in\Lambda'.
\]
\qed
\section{Attachment. Weak solutions of the ITE problem.} We will continue to assume infinite differentiability of $\partial \mathcal O$ and $n$ although a finite smoothness is enough for all the results of this paper. We will call $(u,v)$ a weak interior transmission eigenfunction if it satisfies relations (\ref{Anone0}),(\ref{AnoneB}),(\ref{Antwo}) with weakened assumptions on smoothness of $u$ and $v$. One option is to assume that
$$
u\in L_2(\mathcal O),~v\in L_2(\mathcal O\backslash\mathcal{V}),~u-v\in H^2(\mathcal O\backslash\mathcal{V}).
$$
The last inclusion allows one to define the Dirichlet and Neumann boundary values for $u-v$ on $\partial \mathcal O$ and therefore there are no difficulties in making the boundary condition (\ref{Antwo}) meaningful. In fact, one does not need to assume that $u-v\in H^2(\mathcal O)$. Of course, one can't define the trace at the boundary for arbitrary square integrable functions. However, the traces $u,v\in H^{-1/2}(\partial\mathcal O),~u'_\nu,v'_\nu\in H^{-3/2}(\partial\mathcal O)$ are well defined \cite{roit} for functions $u,v\in L_2(\mathcal O)\times L_2(\mathcal O\backslash\mathcal{V})$ if the latter functions satisfy equations (\ref{Anone0}),(\ref{AnoneB}) (or other elliptic equations). The traces are understood in this case as limits of corresponding traces for smooth approximations of the solutions $u,v$. Thus relations (\ref{Antwo}) for weak interior transmission eigenfunctions  $(u,v)\in L_2(\mathcal O)\times L_2(\mathcal O\backslash\mathcal{V})$
are understood as equalities in the spaces $H^{-1/2}(\partial\mathcal O), H^{-3/2}(\partial\mathcal O)$, respectively.
\begin{theorem}
Let condition (\ref{grcond}) or (\ref{grcond1}) hold. Then $u,v\in H^2(\mathcal O)$ for any weak solutions $(u,v)\in L_2(\mathcal O)\times L_2(\mathcal O\backslash\mathcal{V})$ of the equations (\ref{Anone0}),(\ref{AnoneB}),(\ref{Antwo}), i.e., the set of weak ITE-s coincides with the set of ITE-s in the strong sense defined earlier.
\end{theorem}
\noindent\textbf{Proof.} First assume that a weak ITE $\lambda=\lambda_0$ is neither a pole of $F(\lambda)$ nor a pole of $F_n(\lambda)$. Let $(u,v)\in L_2\times L_2$ be the corresponding weak interior transmission eigenfunction. If $\varphi:=u=v\in H^{-1/2}(\partial\mathcal O)$ is the value of $u=v$ at the boundary, then from (\ref{Antwo}) it follows that $[F(\lambda_0)-F_n(\lambda_0)]\varphi=0,$ i.e., $\varphi$ belongs to the kernel of the elliptic (see Lemma \ref{lemmasigma}) p.d.o. on $\partial\mathcal O$, and therefore $\varphi\in C^\infty(\partial\mathcal O)$. Then $L_2$-solutions $u,v$ of the Dirichlet problems for equations  (\ref{Anone0}),(\ref{AnoneB}) with the Dirichlet data $\varphi\in C^\infty(\partial\mathcal O)$ are infinitely smooth (see \cite{roit}).

Assume now that a weak ITE $\lambda=\lambda_0$ is a pole of $F(\lambda)$ or $F_n(\lambda)$ (or both of these operators). Then from (\ref{Antwo}) it follows that $\varphi=u|_{\partial\mathcal O}=v|_{\partial\mathcal O}$ belongs to the kernel of both the residue of $[F(\lambda)-F_n(\lambda)]$ at the pole $\lambda=\lambda_0$ and the principal part of $[F(\lambda)-F_n(\lambda)]$ at the pole $\lambda=\lambda_0$. The latter property again implies that $\varphi\in C^\infty(\partial\mathcal O)$ since the principal part is also an elliptic p.d.o. on  $\partial\mathcal O$, see Lemma \ref{lemmasigma}. Hence $u,v\in C^\infty$ in all cases.

\qed

\textbf{Acknowledgment.} We are very grateful to the referees for their remarks, which improved the paper significantly and allowed us to avoid some mistakes.

\end{document}